\documentclass[aps,prb,ams,amsmath,twocolumn,reprint,longbibliography]{revtex4-2}
\usepackage{graphicx}
\usepackage{graphics}
\usepackage{epsfig}
\usepackage{amsmath}
\usepackage{amssymb}
\usepackage{amsfonts}
\usepackage{color}

\usepackage{hyperref}
\usepackage{dcolumn}
\usepackage{bm}

\usepackage{siunitx}
\usepackage{makeidx}
\usepackage[italicdiff]{physics}
\usepackage{physics}
\usepackage{mathtools}
\usepackage{hyperref}
\usepackage{cleveref}	
\usepackage{float}	

\usepackage{wrapfig}
\usepackage{ulem}
\usepackage{fancybox,framed}
\usepackage{etoolbox}
\usepackage{changepage}

\begin{document}

\title{Nonreciprocal Coulomb Drag between Quantum Wires in the quasi-1D regime}
\author{R. Makaju}
    \affiliation{Department of Physics, University of Florida, Gainesville, FL 32611, USA}
\author{H. Kassar}
    \affiliation{Department of Physics, University of Florida, Gainesville, FL 32611, USA}
\author{S. M. Daloglu}
    \affiliation{Department of Physics, University of Florida, Gainesville, FL 32611, USA}
\author{A. Huynh}
    \affiliation{Department of Physics, University of Florida, Gainesville, FL 32611, USA}

\author{A. Levchenko}
    \affiliation{Department of Physics, University of Wisconsin Madison, Madison, WI 53706, USA}
    
\author{S. J. Addamane}
    \affiliation{Center for Integrated Nanotechnologies, Sandia National Laboratories, Albuquerque, NM 87185, USA}
\author{D. Laroche}
    \altaffiliation{\textbf{Email of Author to whom correspondence should be addressed:} dlaroc10@ufl.edu}
    \affiliation{Department of Physics, University of Florida, Gainesville, FL 32611, USA}

\date{\today}

\begin{abstract}
Coulomb drag experiments have been an essential tool to study strongly interacting low-dimensional systems.  Historically, this effect has been explained in terms of momentum transfer between electrons in the active and the passive layer. Here, we report Coulomb drag measurements between laterally coupled GaAs/AlGaAs quantum wires in the multiple 1D sub-band regime that break Onsager's reciprocity upon both layer and current direction reversal, in contrast to prior 1D Coulomb drag results. The drag signal shows nonlinear current-voltage (I-V) characteristics, which are well characterized by a third-order polynomial fit. These findings are qualitatively consistent with a rectified drag signal induced by charge fluctuations. However, the nonmonotonic temperature dependence of this drag signal suggests that strong electron-electron interactions, expected within the Tomonaga-Luttinger liquid framework, remain important and standard interaction models are insufficient to capture the qualitative nature of rectified 1D Coulomb drag.
\end{abstract}

\maketitle


\section{Introduction}
Since their first experimental realization nearly four decades ago \cite{van_wees_quantized_1988, thornton_one-dimensional_1986}, one-dimensional (1D) systems have been extensively studied, both to deepen our understanding of strongly correlated systems and for novel quantum applications such as charge sensing \cite{dicarlo_2004}, proximity-induced superconductivity \cite{Doh_2005}, and qubit engineering \cite{mourik_signatures_2012, de_lange_realization_2015, larsen_2015}. In one dimension, the strong confinement leads to reduced screening and increased electron-electron (e-e) interactions \cite{zhang_screening_2009}, giving rise to unique transport phenomena such as interaction-dependent universal scaling \cite{Tarucha_1995}, spin-charge separation \cite{auslaender_spin-charge_2005, jompol_probing_2009}, and charge fractionalization \cite{steinberg_charge_2008}. These seminal experimental results are well understood within the Tomonaga-Luttinger Liquid theory \cite{Tomonaga_1950, luttinger_1963}, where the low-energy excitations of one-dimensional systems are best described by collective spin and charge modes.   

While transport in single quantum wires has been heavily studied experimentally, these experiments did little to deepen our understanding of 1D electron interactions, as the simple conductance measurement in clean systems is expected to yield the noninteracting quantized value \cite{Maslov_1995}, shadowing potential signatures of non-Fermi liquid physics. Instead, experiments between coupled 1D systems have yielded the bulk of the experimental observations of Luttinger liquid physics in 1D systems \cite{auslaender_spin-charge_2005, laroche_1d-1d_2014}. Owing to its sensitivity to both inter- and intrawire e-e interactions, Coulomb drag (CD) \cite{narozhny_coulomb_2016} is one of the prime experimental techniques to study these strongly interacting systems. In a typical CD experiment, a current $(I_{drive})$, sourced in one wire called the drive wire, induces a voltage $(V_{drag})$ in the adjacent drag wire due to e-e interactions, provided that no current is flowing in said drag wire.

Historically, most CD measurements have been interpreted in terms of momentum transfer, owing to their compliance to the Onsager's reciprocity relations \cite{onsager_reciprocal_1931}, as demonstrated in both 2D systems \cite{gramila_mutual_1991, rojo_electron-drag_1999, pillarisetty_frictional_2002, sivan_coupled_1992, seamons_coulomb_2009, kim_coulomb_2011, Gorbachev_2012} and closely separated 1D systems \cite{debray_experimental_2001, Yamamoto_2006, laroche_positive_2011, laroche_1d-1d_2014}. However, recent experiments have reported Coulomb drag signals inconsistent with the simple momentum transfer model, either owing to an unexpected polarity of the drag signal \cite{du_coulomb_2021} or to an explicit breaking of Onsager's relations \cite{tabatabaei_andreev-coulomb_2020,anderson_coulomb_2021, tang_frictional_2020, Guo_2022}. These latter discrepancies are consistent with recent theories \cite{levchenko_coulomb_2008, sanchez_mesoscopic_2010, kaasbjerg_correlated_2016, keller_cotunneling_2016, berdanier_energy_2019} proposing that, in mesoscopic structures, alternate drag-inducing mechanisms involving rectification of charge fluctuations could explicitly break Onsager's relations. Understanding the material and parametric considerations behind the onset of this alternate drag-inducing mechanism is crucial for future developments in the field of coupled 1D systems. 

\begin{figure*}[t!]
\includegraphics[width=\textwidth]{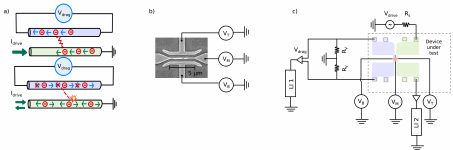} 
\caption{Schematics and circuit diagram of the laterally coupled quantum wires. (a) Schematic representation of the drag-inducing mechanisms due to momentum transfer (top) and energy rectification (bottom). The top wire (blue) is the drag wire and the bottom wire (green) is the drive wire. (b) Scanning electron microscope image of the laterally coupled quantum wires, constituted of a top ($V_T$), a middle ($V_M$) and a bottom ($V_B$) gate. (c) Circuit diagram for Coulomb drag measurements. A drive current is supplied to the green section of the device and the drag voltage is measured in the adjacent wire. The drive current is sourced using a $R_s=10M\Omega$ resistor in series with the drive wire. A virtual ground setup is used on the drag side of the experiment.}
\label{fig:fig1}
\end{figure*} 

In this work, we report CD measured between laterally-coupled quantum wires. In contrast with past studies focusing on the single 1D subband regime that can be understood within the conventional momentum-transfer framework \cite{debray_experimental_2001,laroche_1d-1d_2014} [see upper panel of Fig. 1(a)], we explore the multiple subbands regime at large ($d \gtrsim 150$ nm) interwire separation, where charge rectification has been found to play a predominant role. The reported drag signal shows a clear departure from Onsager's relation and exhibits nonlinear current-voltage characteristics. However its nonmonotonic temperature dependence departs from the expected quadratic dependence predicted in mesoscopic systems with negligible e-e interactions \cite{levchenko_coulomb_2008}, highlighting the likely role that interactions still play within the rectification framework. In the rectification model, depicted in the bottom  panel of Fig. 1(a), the violation of Onsager's relations can be understood by the drive layer creating energy excitations that induce bidirectional momentum transfer in the adjacent layer. However, the wire's energy dependent electron-hole (e-h) asymmetry, intrinsic to mesoscopic devices, results in a drag voltage that is primarily generated in a specific direction, independently of the sign of the drive current. Characterizing this novel drag-inducing mechanism might prove crucial for the development of quantum devices harnessing e-e interactions, particularly in the fields of thermo-electricity \cite{Sothmann_2014, Bhandari_2018} and quantum computing \cite{Schrade_2017}.

The paper is organized as follows: In Sec. \ref{sec:fab} we describe device fabrication and measurement technique. The central section of our work is Sec. \ref{sec:nonreciprocity} where we present key results for the observed nonreciprocity of the drag response. In Sec. \ref{sec:T-dependence} we highlight an anomalous temperature dependence of the drag resistance, which can't be explained based on the conventional paradigm of the Luttinger liquid theory. This lead us to invoke the third-order drag originated from the multiparticle interwire scattering processes as a possible mechanism that could explain the data. We close in Sec. \ref{sec:summary} with a brief discussion and summary of main results.     

\section{Device fabrication}\label{sec:fab}
The coupled quantum wires are fabricated from a GaAs/AlGaAs heterostructure with a quantum well buried $\sim 80\, nm$ below the surface. The quantum wires are laterally coupled over a length $l=5\, \mu m$ and are separated by an electrostatic barrier of width $d\sim 150\, nm$. A scanning electron microscope image of a typical device is shown in Fig. 1(b). The wires are engineered using standard nanofabrication procedures, consisting of both electron-beam and photolithography, and are contacted with evaporated Ge-Au-Ni-Au ohmic contacts. Additional details concerning the fabrication can be found in the Supplemental Material \cite{SM}.

The coupled quantum wires are defined by three gates: a top gate ($V_{T}$), a middle gate ($V_{M}$) and a bottom gate ($V_{B}$) [see Fig. 1(b)]. Unless otherwise specified, standard low frequency lock-in techniques, at a frequency of either $9\, Hz$ or $37.3\, Hz$, are used for the CD measurements.  Additional standard DC measurements have also been performed.  Measurements have been performed in a Bluefors dilution refrigerator, with a base lattice temperature of $\sim 10\, mK$. A circuit diagram of the CD measurement scheme utilizing a virtual ground on the drag side is presented in Fig. 1(c), where $I_{drive}$ is applied on the drive wire (green) and $V_{drag}$ is measured in the drag wire (blue). 

\begin{figure*}[t!]
\includegraphics[width=\textwidth]{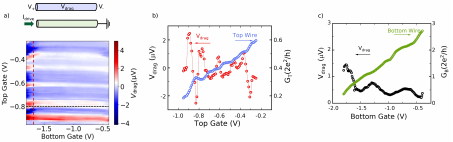} 
\caption{Characterization of the quantum wires.
(a) Drag voltage as a function of top (drive) and bottom (drag) gate voltages. Interwire crosstalk is only visible on the lower end of the top wire gating range. The vertical (red) and horizontal (black) dashed lines represent the line cuts used for panels b) and c) taken at $V_B=-1.7V$ and $V_T=-0.8V$ respectively. (b,c) Drag voltage along their respective line-cut in the top and bottom wires. The conductance plateaus are not quantized at integer values of $2e^2/h$ as the wires are non-ballistic.}
\label{fig:fig2}
\end{figure*} 

A typical CD measurement, over a wide range of subband occupancy in both wires, is shown in Fig. 2(a) while the conductance of both the top and the bottom wire is shown in Fig. 2(b) and Fig. 2(c) respectively, along with a linecut of the drag voltages. The plateaus observed in the conductance of both wires do not lie at integer values of $2e^2/h$ even after accounting for series resistances in the setup, indicating the non-ballistic nature of the wires. The drag signal shows pronounced oscillations over both positive and negative polarities of drag voltage for a given drive current direction, and the oscillations are generally concomitant with openings of 1D subbands in either the drag or the drive wire. As seen from the comparison of the drag peaks in Fig. 2(b) and 2(c), the modulation from the bottom (drive) wire is notably weaker than the one of the top (drag) wire, especially away from the single 1D subband regime. All drag measurements were performed with $V_{M} = 0.15$V, yielding a tunneling resistance larger than $30\, M\Omega$. The drag signal is also frequency independent between $9$ and $85\, Hz$ (see Fig S4).


\section{Nonreciprocal Coulomb drag}\label{sec:nonreciprocity}

To further investigate the discrepancy in the modulation of the drag signal between the top and bottom wires, we measured CD upon layer reversal. Fig. 3(a) shows the CD signal with the bottom wire as the drive wire and Fig. 3(b) with the top wire as the drive wire. The oscillations observed in the drag signal are primarily correlated with the drag wire gate and not strongly correlated with the drive wire gate, as seen from the presence of the horizontal stripes in Fig. 3(a) and vertical stripes in Fig. 3(b). This is a clear violation of Onsager's reciprocity \cite{onsager_reciprocal_1931},  which is expected to be satisfied within the conventional momentum transfer approach to CD. A similar violation occurs upon current direction reversal, as shown in Fig. 3(c) and 3(d). As the current direction is inverted without exchanging the drag voltage probes, Onsager's reciprocity would result in a sign reversal of the drag signal, whereas our measured signal showed minimal changes. These changes, observed when extracting the symmetric and anti-symmetric contributions to the drag signal (see Fig. S5), are less than $\sim 20 \%$ of the symmetric signal, and exhibit reduced modulation with 1D subband occupancy.  These results strongly suggest that conventional momentum-transfer models for 1D CD are inadequate to explain our data.  

\begin{figure}
\includegraphics[width=1\columnwidth]{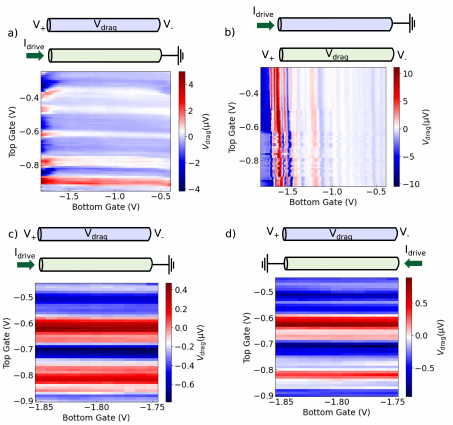} 
\caption{Onsager's relations of the drag signal. The drag voltage is plotted as a function of both top and bottom gate voltages for various measurement configurations. (a) The top wire is used as the drag wire while the bottom wire is used as the drive wire. (b) The top wire is used as the drive wire while the bottom wire is used as the drag wire.  Onsager's relation is not obeyed when the drag and the drive wires are exchanged as the signals are not identical. (c) Same setup as a), but over a different cooldown. (d) The position of current injection in the wire is reversed.  Onsager's relation is broken yet again as the signal's polarity remains virtually unchanged when the current direction is reversed.}\label{fig:fig3}
\end{figure}

\begin{figure}
\includegraphics[width=1\columnwidth]{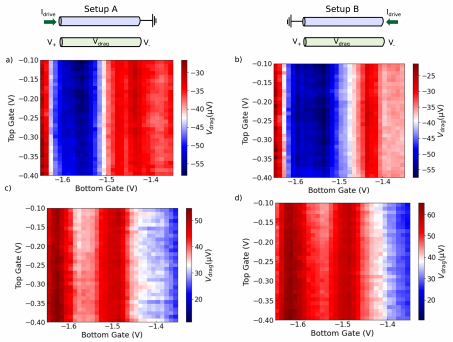}
\caption{Current dependence of the DC drag signal with the top wire as the drive wire and the bottom wire as the drag wire.
(a) DC drag as a function of top and bottom gate voltages with right-flowing $I_{drive}\sim 10$ nA.(b) Same measurement but with the current source position reversed.(c) DC drag as a function of top and bottom gate voltages for right-flowing current $I_{drive}\sim -10$ nA.(d) Same measurement as in c), but with left-flowing current.}
\label{fig:fig4}
\end{figure}

An alternate drag mechanism explaining the violation of Onsager's relationships in mesoscopic systems is due to rectification \cite{levchenko_coulomb_2008} (Fig. 1(a), bottom panel). This model predicts that strong asymmetry, either in e-h transmission probability or the circuit itself, could induce a rectified CD signal that is independent on the drag current direction. A model for rectified CD in coupled Quantum Point Contacts (QPCs) predicts two dominant contributions to drag: a linear contribution from near-equilibrium thermal noise rectification due to e-h asymmetry and a nonlinear contribution, dominating at larger drive currents due to the rectification of quantum shot noise which is sensitive to the circuit intrinsic asymmetry. Both terms are predicted \cite{levchenko_coulomb_2008} to provide the following contribution to the drag signal:

\begin{equation} \label{eqn1}
    I_D^{th} =V\frac{R^2_Q}{4\pi}\int d\omega \frac{\alpha_+(\omega)}{\omega^2}\frac{\partial}{\partial \omega}\left [\coth\frac{\omega}{2T} \right ]\Gamma_1(\omega)\Gamma_2(\omega),
\end{equation}

\begin{equation} \label{eqn2}
    I_D^{shot}=\frac{eV^2}{\Delta_2 R_Q}\alpha_-(0)\sum_{n}|\textbf{t}_n|^2[1-|\textbf{t}_n|^2],
\end{equation}

\begin{equation} \label{eqn3}
    \alpha_{\pm}(\omega) = \frac{Z_{1}^{2}}{8R^{2}_{Q}}
    \frac{C^{2}_{C}}{C^{2}_{L}C^{2}_{R}} 
    \left\{\begin{array}{c}
    2C^{2}_{L} + 2C_{L}C_{R} + 2C^{2}_{R}\\
    C^{2}_{L} - C^{2}_{R}
    \end{array}\right.
\end{equation}
Here, $R_Q=\frac{2\pi \hbar}{e^2}$ is the quantum resistance, $\omega$ is the frequency of the rectified noise from the drive circuit, $\Delta_{2}$ is the energy scale of the confinement potential of the drag wire, $\alpha_{\pm}(\omega)$ is a dimensionless trans-impedance kernel that captures circuitry of interwire interactions approximated in Eq. \eqref{eqn3}. $Z_1$ is the load impedance of the drive circuit, $C_{C}$ is the mutual capacitance between both quantum wires, $C_{L} (C_{R})$ is the capacitance of the left (right) side of the drag wire to ground. Finally, $\Gamma_{1(2)}$ are the rectification coefficients of the drive (drag) wire, given by:
\begin{equation}
\Gamma=\frac{2e}{R_Q}\sum_n\! \int\!d\epsilon[f(\epsilon_-)-f(\epsilon_+)][|\mathbf{t}_n(\epsilon_+)|^2-|\mathbf{t}_n(\epsilon_-)|^2]
\end{equation}
where $\epsilon_-(\epsilon_+)$ is the energy of the electrons (holes) with the corresponding occupations $f(\epsilon_\pm)$, and $\mathbf{t}_n$ is the transmission probability across the $n^{th}$ channel of the wire. Higher order effects can also contribute additional nonlinear terms to the drag signal \cite{borin_coulomb_2019}. We note that, in the linear regime, an e-h asymmetry is essential for the onset of a drag signal, and its sign will depend on whether the carriers transmission probability is locally increasing or decreasing with energy. Within this framework, the left-right Onsager's relation is explicitly broken through the current rectification. In addition, owing to the finite bias across the drive wire, its chemical potential is between $\sim 60$ and $\sim 200$ $\mu$eV larger than that of the drag wire, assuming a drive current of 10 nA and a drive wire conductance ranging from 0.65 to $2.25 \times 2e^{2}/h$. The lack of layer inversion symmetry implies that $\Gamma_1(\omega, \epsilon+60 \mu eV)\Gamma_2(\omega,\epsilon) \neq \Gamma_2(\omega, \epsilon+60 \mu eV)\Gamma_1(\omega,\epsilon)$, \textit{i.e.} that the wire's rectification coefficients are not identical. 

\begin{figure}
\includegraphics[width=1\columnwidth]{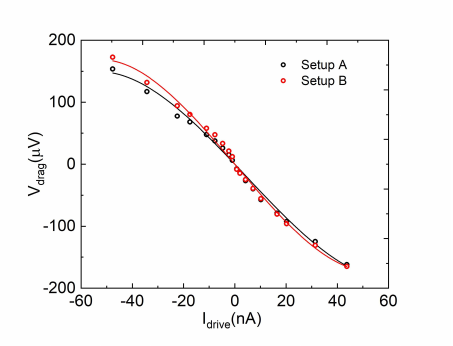}
\caption{Nonlinearity of the DC drag signal.
The IV relationship of the drag signal is presented for both current directions (setups A and B) and for both polarity of the drive current at a top gate voltage of -1.51V and a bottom gate voltage of -0.284V. The data is well fitted by third order polynomial function.
}
\label{fig:fig5}
\vspace{-7mm}
\end{figure}

\begin{figure*}[t!]
\includegraphics[width=\textwidth]{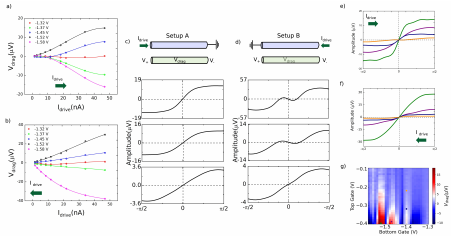} 
\caption{Waveform and I-V characteristics of the drag signal in two different setups. 
(a) Drag voltage as a function of drive current, for setup A, as shown in the top of panel c).
(b) Same plot for setup B, with inverted current direction. The solid lines represent a cubic fit. The I-V relationship deviates from the linear behavior predicted by momentum transfer models.
(c) The waveforms of the drag signal at different drive currents: $1.46$ nA, $5.38$ nA and $18.34$ nA (from bottom to top) for wire setup A.
(d) Same plot as (d) for setup B. The peak of the waveform increases in magnitude as the drive voltage is increased from $1.46$ nA to $18.34 $ nA.
(e) The waveform of the drag signal at different gate voltages for setup A.
(f) Same waveform plot as (e) for setup B.
(g) Drag voltage as a function of top gate and bottom gate voltages. The colored points on the plot corresponds to the waveform plots in (e) and (f).}
\label{fig:fig6}
\end{figure*}

Studying the DC response of the drag signal simplifies the measurement by fixing the electrons chemical potential to a single value in the drive wire.  As presented in Fig. 4, we measured the DC 1D drag with the top wire as drive wire and the bottom wire as the drag wire, in both current directions and with both positive [Fig. 4(a) and (b)] and negative current sources [Fig. 4(c) and (d)]. As in the AC drag, the DC drag violates Onsager's relation upon reversal of the current direction.  However, the signal changes both in magnitude and sign by going from positive to negative voltages. This further corroborates the rectified CD model, as only the chemical potential of the drive wire has an incidence on the drag signal, and not the direction of the current flow. As expected for rectified drag dominated by the linear component, the sign of the drag voltage also inverts when the sign of the drive voltage is inverted.  To quantify the nonlinearity of the drag signal, we present in Fig. 5 the I-V relationship of the DC drag voltage.  In setup A (see Fig. 4), the drag voltage is well described by $V_{drag} = -5.5 I + 0.058 I^{2} - 3.4 \times 10^{-4} I^{3}$, with the current given in nA and the voltage in $\mu$V.  A cubic fit was selected, as neither a linear fit nor a quadratic fit provided a good match to the data.  

\begin{figure*}[t!]
\includegraphics[width=\textwidth]{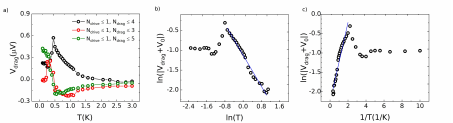} 
\caption{Temperature dependence of the CD signal. (a) Temperature dependence of the CD signal with $N_{drive} \le 1$ and $N_{drag}<=4$ (black), $N_{drive} \le 1$ and $N_{drag}\le 3$ (red) and $N_{drive} \le 1$ and $N_{drag}\le 5$ (green). 
(b) Log-log plot of drag voltage and temperature for $N_{drive} \le 1$ and $N_{drag}\le 4$. The blue straight line is the linear fit for high temperature regime and the offset is $V_0 = 0.16451 \mu V$. 
(c) Arrhenius plot of  drag voltage and temperature for $N_{drive} \le 1$ and $N_{drag} \le 4$, with a linear fit (blue straight line) in the high temperature regime. The offset is  $V_0 = 0.16451 \mu V$.}
\label{fig:fig7}
\end{figure*} 

A similar nonlinearity of the drag I-V relation is observed in the AC regime, as shown in Figs. 6(a) and (b). Consistently with the microscopic model for rectified drag in mesoscopic circuits, quantitative details of the drag nonlinearity strongly vary with gate voltage. Over all gate voltages analysed, the linear coefficients (in $\mu V/nA$) are between 1 and 2 order of magnitudes stronger than the quadratic terms (in $\mu V/nA^{2}$), which are themselves between 1 and 2 orders of magnitude larger than the cubic terms (in $\mu V/nA^{3}$) (see Tables S3 and S4 for parameters details). As such, the predominant contribution to the drag signal appears to be rectification of near-equilibrium thermal-noise, but quantum-shot noise rectification is still significant for certain gate voltage configurations.  The discrepancy in the fitting parameters between the DC and the AC measurements can be explained by microscopic changes in the wires' potential landscape between different cooldowns.
Indeed, in the mesoscopic regime, one would naturally expect that the magnitude of CD fluctuations is controlled by the Thouless energy, $E_{\text{Th}} = \frac{\hbar D}{L^{2}} = \frac{\hbar v_{F}l}{2 L^{2}}$ \cite{Aleiner-PRL2000, Price-Science2007}.  Here, $D$ is the mesoscopic system diffusion coefficient, $L$ is the wire's length, $l$ is the electron mean free path, $v_F=\frac{\pi \hbar n_{1D}}{2m^*}$ is the Fermi velocity, $n_{1D}$ is the electronic density and $m^{*}$ is the electron effective mass. 

Estimating our 1D electron density from magnetic depopulation measurements \cite{Berggren_1988} (see Supplemental Material \cite{SM}), a 1D density of $n_{1D} \sim {8.94\times 10^{8}\, m^{-1}}$ is estimated when the wire has 5 populated subbands, and $n_{1D} \sim {1.06 \times 10^{9}\, m^{-1}}$ when the wire has 6 populated subbands. The mean free path can be estimated from the typical size at which quantized 1D conduction is observed, which is $\sim 1 \, \mu$m in shallow 2DEGs. From these estimates, we obtain a Thouless energy in the range of $E_{\text{Th}} \sim 6 - 10 \, \mu$eV, in good agreement with the typical size of the oscillations observed in our CD signal.   

The nonlinearity of the current-voltage characteristics should also result in a deformation of the sinusoidal drag signal in AC measurements. Fig. 6(c) and (d) show the waveforms of the drag signal from $-\pi /2$ to $-\pi /2$ at different drive currents (1.46 nA, 5.38 nA and 18.34 nA from bottom to top) for wire setups A [Fig 6(c)] and B [Fig 6(d)] respectively, with bottom gate at -1.52 V and top gate at -0.38 V.  These waveforms were calculated by adding the first 9 harmonics of the drag signal. The shape of the waveform is drive current dependent and its magnitude increases as the drive current is increased. Fig. 6(e) and (f) shows the waveforms for $I_{drive}=11.8\, nA$, at different top and bottom gate voltages which are represented by the filled circles in Fig. 6(g). Due to the nonlinear I-V relation, we expect the waveforms to digress from the expected sinusoidal shape, and exhibit significant dependence on values
of the gate voltages. The waveforms display nonidentical characteristic upon current direction reversal, likely caused by a small momentum-transfer contribution to the drag signal or different line resistances into the wires, changing the electron's chemical potential.  We also note that Joule heating from the drive current ($\sim 1 mV$ voltage drop at 10 nA) is unlikely to be at the origin of the I-V nonlinearity, since, as shown in Fig. 7, the drag signal resulting from a 10 nA drive current exhibits a nonmonotonic temperature dependence down to $\sim 180$ mK, a much lower temperature than the voltage temperature of the drive circuit ($\sim 1.6$ K).


\section{Temperature dependence}\label{sec:T-dependence}

In the Fermi liquid regime, CD induced from charge density fluctuations is expected to depend quadratically on the temperature.  However, as presented in Fig. 7(a), the observed temperature dependence of the drag signal is nonmonotonic. The observation of both an increasing drag signal  with a decreasing temperature \cite{fuchs_coulomb_2005} and of a nonmonotonic temperature dependence \cite{pustilnik_coulomb_2003, narozhny_coulomb_2016} are hallmarks of interaction effects within the Luttinger liquid model, albeit in a framework where the drag signal is induced by momentum transfer.  However, to describe this effect in the diffusive limit of multichannel quantum wires, one must go beyond the usual approximations of the Fermi liquid and Luttinger liquid theories of drag. In particular, the three-particle interwire correlations remove the constraints of particle-hole asymmetry and may lead to a strong drag effect in the low-temperature regime. In the diffusive limit, $T\tau \ll 1$, where $\tau$ being the intrawire transport scattering time, the resulting temperature dependence of the third-order drag mechanism of transconductivity can be extracted from Ref. \cite{levchenko_third-order_2008} with modifications appropriate for the 1D system. We find $\sigma_{D} \sim R^{-1}_Q (\nu U_0)^{3} L_{T} \propto 1/T$ for the case of short-ranged interactions (strong screening), where $L_{T} = \frac{v_{F}}{T}$ is the thermal de Broglie lengths, and $\nu$ is the 1D-density of states, and $U_0$ is the characteristic strength of the interwire interaction for forward scattering with small momentum transfer. The surprising feature of this result is that it is independent of $\tau$, both for the temperature dependence and its prefactor. For long-ranged interactions, we find the same temperature dependence, but with a more rapid decay of drag with the interwire separation, namely $\sigma_{D} \sim R^{-1}_Q (\nu U_0)^{3} L_{T}/(\kappa d)^{3}$ for $\kappa d\gg 1$ where $\kappa$ is the inverse Thomas-Fermi screening radius. An extension of the formalism from Refs. \cite{levchenko_third-order_2008, mora_2007} to the ballistic limit of transport $T\tau>1$ results in the Fermi-Liquid like temperature dependence of drag conductivity $\sigma_{D} \sim R^{-1}_Q (\nu U_0)^{3} (v_{F} \tau) (T\tau)^{2} \propto T^{2}$. Therefore, that three-particle mechanism of drag can result in both a nonmonotonic temperature-dependence and an upturn of drag at low temperatures, even from the forward electron scattering at small momentum transfer between the wires. This analysis should be contrasted to the Arrhenius behavior predicted to occur in ballistic wires for interwire backscattering between strongly correlated wires \cite{fuchs_coulomb_2005} and for interwire forward scattering \cite{dmitriev_coulomb_2012, dmitriev_ultranarrow_2016}. 

We present the result of power-law fits of the drag temperature dependence in Fig. 7(b) and (c), in log-log and Arrhenius form respectively. The blue solid line in Fig. 7(b) and 7(c) indicates the regime where the log-log plot and the Arrhenius plot is nearly linear, and the exponent for these fittings were calculated as $V_{drag} \propto T^{\gamma}; \gamma = -0.98 \pm 0.04$ for the power-law function and $V_{drag} \propto e^{\frac{\beta}{T}}; \beta = 1.028 \pm 0.01$ for the Arrhenius function. Analysis at different subband occupancies lead to a comparable power-law fit:  $V_{drag} \propto T^{\gamma}; \gamma = -0.8 \pm 0.2$. We note that, owning to a sign change at high temperature in our data, an offset voltage $V_{0}$ has been included in the fit. Additional details about the fitting procedure can be found in the Supplemental Material \cite{SM}. While the power-law exponent value is consistent with the three-particle mechanism for Coulomb drag described prior, the limited range where the drag signal is showing an increase with decreasing temperature prevents us from ruling out the possibility that a more conventional Arrehenius-like behavior is occurring. Additional experimental and theoretical work will be required to confirm this conclusion.  


\section{Discussion and summary}\label{sec:summary}

The results reported in this paper are fairly different from prior 1D drag results \cite{debray_coulomb_2002, Yamamoto_2006, laroche_1d-1d_2014} where the CD signal appeared to be consistent with the momentum transfer model. The reason behind this discrepency is not readily apparent. However it is likely that a combination of the large subband occupancy in the wires, the significant interwire separation and the sample innate disorder could  be the source of these fundamental differences in the nature of the dominant drag-inducing mechanism. It should also be noted that, as highlighted by recent studies \cite{anderson_coulomb_2021, song_coulomb_2013, tang_frictional_2020, du_coulomb_2021}, observations of a negative and/or nonreciprocal CD is not uncommon in mesoscopic systems.  Additional experimental and theoretical work will be required to determine the universality of rectification-induced drag across various material platforms and to assess the parametric onset of both momentum transfer and rectification induced drag. 

In summary, we present an experimental study of 1D Coulomb Drag between quantum wires in the multiple subband regime. Our CD measurements deviates from the standard momentum transfer models by clearly violating the Onsager reciprocity relations, both upon layer reversal and current reversal. Subsequent measurements of the nonlinearity of the drag signal are consistent with a microscopic energy rectification model for Coulomb drag. However, the nonmonotonic temperature dependence of the drag signal highlights the importance of including electron-electron interactions beyond the Luttinger liquid framework in future theoretical description of rectification-induced drag.\\

\section*{Data availability}
The supporting data for this article are openly available from the Institutional Repository at the University of Florida (IR@UF) \cite{Data}.

\section*{Acknowledgements}
This work was performed, in part, at the Center for Integrated Nanotechnologies, an Office of Science User Facility operated for the U.S. Department of Energy (DOE) Office of Science. Sandia National Laboratories is a multimission laboratory managed and operated by National Technology \& Engineering Solutions of Sandia, LLC, a wholly owned subsidiary of Honeywell International, Inc., for the U.S. DOE's National Nuclear Security Administration under contract DE-NA-0003525. The views expressed in the article do not necessarily represent the views of the U.S. DOE or the United States Government. This work was partially supported by the National High Magnetic Field Laboratory through the NHMFL User Collaboration Grants Program (UCGP). The National High Magnetic Field Laboratory is supported by the National Science Foundation through NSF/DMR-1644779 and the State of Florida. A. L. acknowledges support by the NSF Grant No. DMR-2203411 and H. I. Romnes Faculty Fellowship provided by the University of Wisconsin-Madison Office of the Vice Chancellor for Research and Graduate Education with funding from the Wisconsin Alumni Research Foundation.

%


\end{document}